# Interaction Between Dilute Water Vapor and Dodecane Thiol Ligated Au Nanoparticles: Hydrated Structure and Pair Potential of Mean Force

Michael N. Martinez[1,a)], Alex G. Smith[1,b)], Linsey M. Nowack[1,c)], Binhua Lin[1,2] and Stuart A. Rice[1,3,4,d)]

[1]James Franck Institute, University of Chicago, 929 E. 57th Street, Chicago, IL 6037
[2] NSF ChemMatCARS, University of Chicago, 929 E. 57th Street, Chicago, IL 6037
[3]Department of Chemistry, University of Chicago, 929 E. 57th Street, Chicago, IL 6037
[4]Chicago Center for Theoretical Chemistry, University of Chicago, 929 E. 57th Street, Chicago, IL 6037

[a)]Current Address: Department of Physics, University of Wisconsin, Madison
[b)]Current Address: Department of Physics, University of California, Berkeley
[c)]Current Address: Department of Chemistry, Massachusetts Institute of Technology
[d)]Author to whom correspondence should be addressed: sarice@uchicago.edu

## Abstract

The interaction between two ligated nanoparticles depends on whether they are isolated or immersed in a liquid solvent. However, very little is known about the influence of solvent vapor on the interaction between two ligated nanoparticles. Recent experiments yield the surprising result that cyclic exposure of solvent free suspended monolayers of dodecane thiol ligated gold nanoparticles (AuNP) to water vapor and dry nitrogen generates reversible cyclic decreases and increases of the Young's modulus of the monolayer, implying corresponding cyclic changes in the AuNP-AuNP interaction. We examine how water vapor interacts with an isolated dodecane thiol dressed AuNP, and how water vapor affects the interaction between a pair of nanoparticles, using all-atom molecular-dynamics simulations. We find that there is condensation of water molecules onto the ligand shell of an AuNP in the form of clusters of 100-2000 molecules that partially cover the shell, with most of the water in a few large clusters. A water cluster bridges the AuNPs, with a sensibly constant number of water molecules for AuNP-AuNP separations from edge-to-edge contact up to center-to-center separations of 100 Å. The wet AuNP-AuNP interaction has a slightly deeper and wider asymmetric well than does the dry interaction, a change that is qualitatively consistent with that implied by the observed water vapor induced change of the Young's modulus of a monolayer of these AuNPs. We find that macroscopic analyses of water drop-deformable surface interactions and dynamics provide both guidance to understanding and qualitatively correct predictions of the phenomena observed in our simulations.

## I. Introduction

Immersion of a long chain alkane ligated nanoparticle in a solvent alters the conformations of the ligand chains in a fashion that depends on the qualitative character of the ligand-solvent interaction.[1-3] In a good solvent the ligands have extended conformations, and in a poor solvent the ligands have compact conformations, with consequences for the pair potential of mean force between the dressed nanoparticles. Specifically, when immersed in a good solvent the pair potential of mean force is, typically, everywhere repulsive whereas when immersed in a poor solvent the pair potential of mean force has a strong minimum.[4-6] These general conclusions follow from the results of a substantial number of molecular dynamics simulation studies of the ligand structure of isolated and immersed dressed nanoparticles, of the interactions between ligated nanoparticles, and of the changes in solvent structure in the vicinity of the ligand chains.[1-7] These studies also establish that, for typical choices of ligands, the interaction between the ligands, not the core-core interaction, determines the pair potential of mean force of the dressed nanoparticles.

Unlike immersion in a liquid solvent, very little is known about the influence of solvent vapor on the interaction between two ligated nanoparticles. Neither the character of solvent-ligated nanoparticle configurations when the nanoparticle is in a low-density solvent vapor, nor the change, if any, in pair potential of mean force between ligated nanoparticles exposed to low-density solvent vapor, have been examined. This lack of information is an understandable consequence of the focus on applications of nanoparticles that are assembled in solvent free two-dimensional and three-dimensional dense aggregate forms, e.g. the mechanical properties of the close-packed monolayer.[1,8] Nevertheless, it is known that under ambient conditions nanoscale water at the molecular level is almost always present on nominally dry surfaces, including hydrophobic surfaces such as formed by monolayers of ligated nanoparticles.[9] Evidence that adsorption of low-density water vapor induces changes in monolayer mechanical properties is provided by a recent experimental study of the change in Young's modulus of a freely suspended self-assembled monolayer of dodecane thiol ($CH_3(CH_2)_{11}SH$) ligated Au nanoparticles (AuNPs), which reveals large magnitude reversible decreases/increases of the Young's modulus of the monolayer generated by cyclic exposure to and removal of 25°C water vapor.[10]

An ordered monolayer of AuNPs is a complicated many-body system, the energy of which is not accurately represented by the sum of pair interactions. Consequently, we must expect there to be an important contribution to its mechanical properties from multi-AuNP interactions.[11] Even without dissection into separate multi-particle contributions to the monolayer energy, the results of the dry/wet Young's modulus experiment strongly imply that exposure to dilute water vapor has altered the AuNP-AuNP pair potential of mean force.[10] It is that inference that stimulated the calculations described in this paper.

The change in pair potential of mean force generated by exposure of two AuNPs to water vapor is intrinsically interesting because the ligand shell of the AuNP is hydrophobic, hence the distribution of water molecules on the shell, and the conformations of the ligands bound to the Au core, will be different from those when the AuNP is immersed in liquid water. Indeed, the adsorption of equilibrium vapor pressure water at 25°C on flat saturated hydrocarbon surfaces has been experimentally determined to amount to about 80% of a monolayer, with the water deposited on the surface in the form of separated water droplets, not as a uniform monolayer. This conclusion is supported by simulations[12-14] of the wetting of self-assembled monolayers of long chain hydrocarbons by water that also show that the water molecules adsorbed on the $-CH_3$ chain ends aggregate to form a compact droplet.

In this study, we report molecular dynamics simulations of the interaction of water vapor with isolated AuNPs with a 50 Å diameter Au core that have 78% and 98% ligand coverage, and the pair potential of mean force between AuNPs with either 78% or 96% ligand coverage. The simulations reveal the role played by the structure of water clusters around an isolated nanoparticle and around a pair of nanoparticles. And, substituting for the complexity and scale of a calculation of the free energy of deformation of an ordered AuNP monolayer, we use the change in pair potential of mean force generated by exposure of two AuNPs to water vapor as a surrogate indicator of how exposure to water vapor changes the Young's modulus of the AuNP monolayer.

## II. Simulation Details

We have simulated various nanoparticle systems using the LAMMPS and DASH molecular dynamics packages (see Table 1).[15,16] The DASH package was used to equilibrate the ligand chains on the Au core when the composite nanoparticle was created, while the LAMMPS package was used to calculate the interaction between the AuNPs. All simulations were conducted at 300 K.

**Table 1.** Molecular dynamics simulations details

| **Simulation type** | **Number of Orientations** | **Ligand Coverage (ligands/nm$^2$)** | **Number of ligands per nanoparticle** | **Fractional Ligand Coverage** | **Simulation box dimensions (Å$^3$)** | **Number of water molecules** | **Water initialization** |
|---|---|---|---|---|---|---|---|
| Single AuNP in water vapor | 1 | 3.6 | 282 | 78% | 200 × 200 × 200 | 2364 | Diffuse lattice filling whole simulation box |
| | 1 | 4.5 | 352 | 98% | 200 × 200 × 200 | 2364 | Diffuse lattice filling whole simulation box |
| Pair of AuNPs in vacuum | 8 | 3.6 | 282 | 78% | 200 × 100 × 100 | 0 | N/A |
| | 3 | 4.4 | 345 | 96% | 200 × 100 × 100 | 0 | N/A |
| Pair of AuNPs in water vapor | 3 | 3.6 | 282 | 78% | 250 × 150 × 150 | 4810 | Diffuse lattice filling whole simulation box |
| | 3 | 4.4 | 345 | 96% | 250 × 150 × 150 | 4810 | Diffuse lattice filling whole simulation box |
| | 1 | 3.6 | 282 | 78% | 250 × 150 × 150 | 4810 | Dense lattice encasing two nanoparticles |

### IIA. AuNP model

Our model of the ligated gold nanoparticle features a 50 Å diameter Au core represented as an impenetrable sphere. Immobile dodecane thiol ligands are uniformly distributed around the spherical core. We generate this distribution by first randomly placing the head groups on the surface of the Au core, then allowing the ligands to move on that surface under the influence of the ligand-ligand interaction until the coverage is on average uniform, at which stage they are frozen in place. At both ligand coverages considered, the head groups and ligand chains on the spherical surface are somewhat disordered. The center of the S atom is placed at the Au core radius and lies half-in, half-out, of the core. Although this representation neglects details of the Au-S bonding arrangement and initial ligand chain tilt, we argue that it is acceptable because the water molecules adsorb on the ends of the ligand chains. The intramolecular and intermolecular interactions of the ligand chains were represented by the OPLS-AA force field modified to account for long alkane chains with the parameter set developed by Siu, Pluhackova and Bockmann.[17]

A number of studies over the last decade suggest that an AuNP monolayer's mechanical properties depend on the fractional coverage of ligands on the nanoparticle core surface.[18,19] Moreover, this fractional surface coverage may be altered via changing the concentration of free thiols to which the AuNP is exposed. Therefore, in our single nanoparticle simulations, we compare the ligand conformations of two surface coverages: 3.6 ligands/nm$^2$ and 4.5 ligands/nm$^2$. Using 4.6 ligands/nm$^2$ as the maximal coverage of ligands on a 50 Å diameter gold core, then the above two coverages are 78% and 98%.[19] At 78% coverage, our model AuNP has 282 ligands distributed around the core surface, and at 98% coverage the Au core carries 352 ligands. The simulations with two AuNPs were conducted at 78% and 96% (4.4 ligands/nm$^2$; 345 ligands), with the difference in the higher-coverage from the single-AuNP case imposed by simulation constraints.

At first sight, choosing to study AuNPs with 78% ligand coverage seems arbitrary and odd. Our rationale for this choice is derived from the simulations of Wang et al.,[10] who compared the measured Young's modulus with that obtained from coarse grained molecular dynamics simulations of a $16 \times 16$ assembly of AuNPs with 78% ligand coverage. We have, therefore, adopted 78% ligand coverage as a link to experiment in our calculation of the influence of water vapor on the pair potential of mean force. The higher ligand coverage simulations correspond to the experimental portion of Wang et al.[10]

All simulations were carried out in the NVT ensemble with periodic boundary conditions at 300 K using a Nosé-Hoover-thermostat and a 1 fs timestep. For single AuNP simulations, a 200 Å × 200 Å × 200 Å simulation box was used. For simulations of a pair of AuNPs in vacuum, the box size was 200 Å × 100 Å × 100 Å, and for those in water vapor, 250 Å × 150 Å × 150 Å. In both of the latter simulations, the interparticle axis was along the (long) x axis. The systems were equilibrated for 1-2 ns, after which the particles were slowly moved towards each other, and the state of the simulation was saved at a predetermined list of core-core separations. These saved states were restarted and equilibrated for another 1-2 ns before data collection. The fluctuations in the equilibrated state have a magnitude of $\pm 0.15\%$ total energy.

**IIB. Water vapor model**

In our simulations we have used the Simple Point-Charge (SPC) model of water, which provides a good but not perfect description of water structure and thermodynamics.[20-22] For the single AuNP in water vapor calculations, the simulation box contained 2364 SPC water molecules that, in the initial state, are uniformly distributed on a lattice in the complementary space outside of the AuNP. This choice of number of water molecules is derived from the study of water adsorption on alkane surfaces reported by Thomas et al.,[23] who found that at 25°C and 100% relative humidity (0.0313 atm vapor pressure) the adsorbed water occupied the equivalent of 80% of a monolayer. We note that in the initial state of the simulation, with no water adsorbed on the AuNP, the water vapor is extremely supersaturated.

For the pair of AuNPs in water vapor simulations, the simulation box contained 4810 water molecules with similar initialization on a diffuse lattice. This number of water molecules also corresponds to 100% relative humidity based on the data of Thomas et al.[23] After equilibration, at any given time, our wet system contains adsorbed water molecules and a small number of free individual water molecules, usually between 3 and 10, consistent with the expected equilibrium number of free water molecules ($\approx$ 4 - 5) in the simulation cell volume. Tracking of the free water molecules reveals that their identities change but their number is maintained, which is consistent with the desired adsorption coverage determined from the water vapor-alkane surface equilibrium.

In addition to calculations whose initial states have gaseous water molecules uniformly distributed in the volume of the simulation cell, we also have carried out simulations in which the initial state has two AuNPs linked by a dense liquid slab that wets both and has the form of a liquid collar surrounding the line of centers. This liquid slab contains all the water molecules, so defines an initial condition very different from that with water molecules dispersed throughout the simulation cell volume. The rationale for this choice of initial condition will become evident in the discussion below of the results of the simulations of the wet AuNP-AuNP pair potential of mean force.

**IIC. Pair potential of Mean Force Calculation**

In both our wet and dry simulations, the interaction between two AuNPs was sampled at discrete center-to-center separations from 100 Å to 58 Å. At each separation we collected data for 2 ns after reaching equilibrium. The thiol head group distribution on the Au core is kept fixed along the separation path between the two AuNPs, whilst the thermal configurations of the ligand chain are equilibrated. To calculate the effect of AuNP rotation, we averaged over nine thiol head group distributions created as new initial states, each of which has a somewhat different set of ligand attachments to the Au core. These different ligand distributions lead to a set of different face-to-face ligand shell orientations that are surrogate for rotation. An impression of the distribution of head group spacing can be obtained from a VMD rendering of an isolated AuNP with water clusters (See Section IIIB below). The interaction between two AuNPs at some specified center-to-center separation is, directly, the potential of mean force by virtue of the averaging over ligand chain conformations and face-to-face orientations. Our calculations yield the total energy of the pair of AuNPs. We represent the interaction relative to a zero of energy at very large separation of the AuNPs by subtraction of the energy at large separation from that at any specified separation. As a matter of interest, we include the range of interaction energy at fixed AuNP-AuNP separation generated by the different face-to-face ligand distributions with bars (Figs. 4, 9 and 10 below). For the wet system, at each new separation, the system was initiated with water molecules distributed on a lattice throughout the simulation cell volume. These calculations allow us to compare the dry pair potentials of mean force between AuNPs with 78% and 96% ligand coverage, the wet pair potentials of mean force between AuNPs with the same coverages, and the differences between the dry and wet pair potentials of mean force. Because of the much greater number of molecules in the computations, the averages over face-to-face orientations in the wet system involved five face-to-face orientations.

We also compare our calculated interaction between all-atom AuNPs in vacuum with the interaction calculated for united-atom AuNPs in vacuum by Liepold et al.[18]

## III. Simulation Results

### IIIA. Fractional Surface Coverage of a Single AuNP.

We show in Section IIIB that the water adsorbed on the ligand shell of an AuNP forms clusters, not a monolayer. We examine first how that form of fractional surface coverage affects the spatial arrangement of ligand carbons for a single AuNP in vacuum and in water vapor. Figure 1 displays the radial distributions of carbon atoms of an isolated AuNP with surface coverages of 3.6 ligands/nm$^2$ and 4.5 ligands/nm$^2$. The density of the ligand brush of the AuNP is about 25% greater when the coverage is 4.5 ligands/nm$^2$ than when it is 3.6 ligands/nm$^2$. Viewed overall, when the ligand density is increased from 3.6 to 4.5 ligands/nm$^2$ the structure in the carbon atom radial distribution close to the attachment to the Au core is slightly sharpened, and the peak in the radial distribution of the last carbon atom is slightly shifted in the direction of full extension of the ligand chain. In the presence of water vapor (see next Section), for both ligand densities, the peak in the last carbon radial distribution function of ligands under a water cluster is shifted to smaller average chain extension.

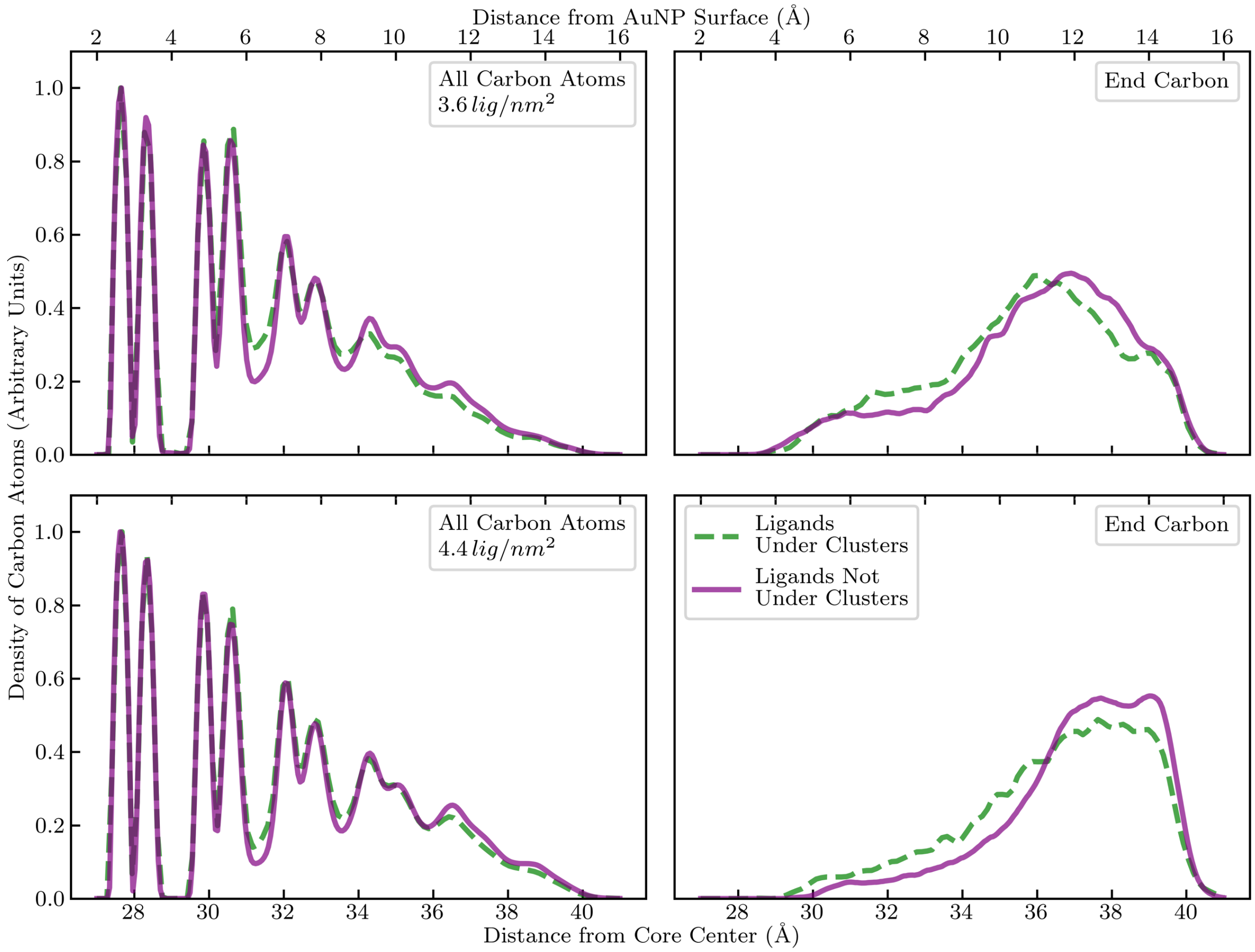

Fig. 1. Comparison of the radial distributions of carbon atoms of an isolated AuNP in water vapor with ligand coverages of 3.6 ligands/nm$^2$ (upper panels) and 4.5 ligands/nm$^2$ (lower panels). The left panel shows the radial distribution of all 12 carbon atoms in the ligand chain, and the right panel shows the radial distribution of the end carbon. The change in end carbon distribution in ligands under clusters further illustrates the flattening effect the cluster has on the ligands. Note that the (arbitrary units) carbon atom density scales in the left and right panels are different. In the left panels the scale refers to all carbons in the ligand chain, whilst in the right panel it is only for the end carbon.

### IIIB. Water Clusters on a Single AuNP in Water Vapor

The most obvious result of our simulations of the molecular configuration in the water vapor-AuNP system is the condensation of water molecules onto the ligand shell in the form of clusters that partially cover the shell. Starting with water molecules distributed throughout the simulation cell, the stationary state achieved on the time scale of our simulations always has a few water clusters on the ligand shell and a very few water molecules in the vapor. With this choice of initial water configuration, we do not observe coalescence of all the water into one cluster. An example set of configurations for the AuNP with coverage of 3.6 ligands/nm$^2$ at different times after equilibration is displayed in Fig. 2. The clusters attached to the ligand shell range in size, with the smallest containing of order 100 water molecules and the largest containing of order 1000 water molecules. The small clusters are more mobile than the large clusters. An initially independent small cluster sometimes coalesces with a large cluster, and sometimes a large cluster ejects a small cluster. We observe only rare coalescences of large clusters and their reverses (Fig. 2). Although they exhibit fluctuations in shape, the area of contact of a large water cluster with the ligand shell shows little variation. This behavior is illustrated in the supplemental video version of Figure 2.

Our simulation results agree with the experimental results of Thomas et al. that show that at 25$^o$C water vapor adsorbed on a smooth close packed surface of a film of $CH_3(CH_2)_{11}SCl_3$ deposited on quartz is condensed in droplets localized at imperfections in the alkane surface.[23] Our results also agree with those obtained in simulation studies of water adsorbed on a flat ordered monolayer of alkane thiols that show the water to form a liquid droplet.[12-14] With respect to the interaction of the water droplet and the hydrocarbon surfaces, we find the large cluster water contact angle to be $120^o \pm 10^o$ with the 78% ligand covered AuNP and $120^o \pm 10^o$ with the 96% ligand covered AuNP. Taking note of small differences in the potentials employed in the several simulations, our results also agree with those of Mar et al. (SPC water) who find the water-$CH_3(CH_2)_{11}S$ monolayer contact angle to be $135^o \pm 10^o$,[12] with those of Mousa et al. (SPC/E) who find the water-$CH_3(CH_2)_7S$ and water-$CH_3(CH_2)_{17}S$ monolayer contact angles to be $107^o \pm 1^o$ and $112^o \pm 2^o$, respectively,[13] and with those of Thomas et al. who report an experimental value of 115$^o$ for the contact angle of water with a smooth close packed surface of a film of $CH_3(CH_2)_{11}SCl_3$ deposited on quartz.[23]

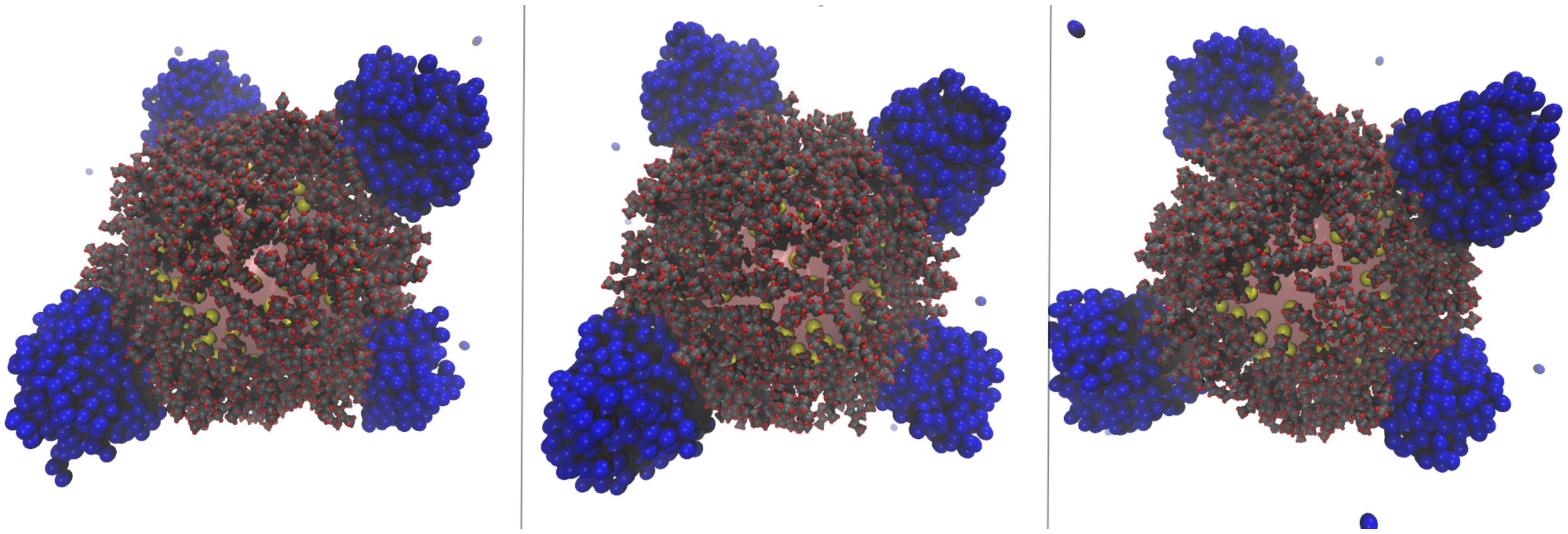

Fig. 2: VMD renditions of the water clusters adsorbed to the ligand shell of the AuNP (3.6 ligands/nm$^2$) at different times (left to right 3.0 ns, 4.0 ns and 5.0 ns) after equilibration.

Given that water molecules adsorb onto the AuNP surface in clusters, not as a monolayer, we now investigate how the part of the ligand shell under a water cluster differs from the part not covered, via examination of a single AuNP in water vapor. We start with the observation that the ligand shell of the AuNP is not rigid, so that the total free energy of the liquid drop-ligand shell interface will include the free energy of elastic deformation of the portion of the ligand shell in contact with liquid. Indeed, minimization of the total surface free energy must include changes of shape and area of the part of the ligand shell in contact with the water droplet. Thus, an analysis of the deformation of the surface regions of two elastic spheres connected with a liquid bridge, as reported by Zheng and Streator,[24] shows that the surface of an elastic sphere in contact with the liquid bridge is deformed (flattened). We show in Fig. 3 images of the layers of water in clusters in contact with the 3.6 ligands/nm$^2$ and the 4.5 ligands/nm$^2$ shells of the AuNPs; the flattening of the ligand shells is evident. This visual inference is supported by an examination of the radial distributions of ligand carbon atoms under and not under water clusters for an isolated AuNP that are displayed in Fig. 1 for the AuNPs with coverage of 3.6 ligands/nm$^2$ and 4.5 ligands/nm$^2$. The data imply that although the distribution of all carbons in the chain (Fig. 1, left panel) is insensitive to the water, the distribution of the end carbon (right panel) corresponds to a densification of the outer region of the ligand shell under the water cluster, consistent with the flattening of the ligand shell as seen in Fig. 3.

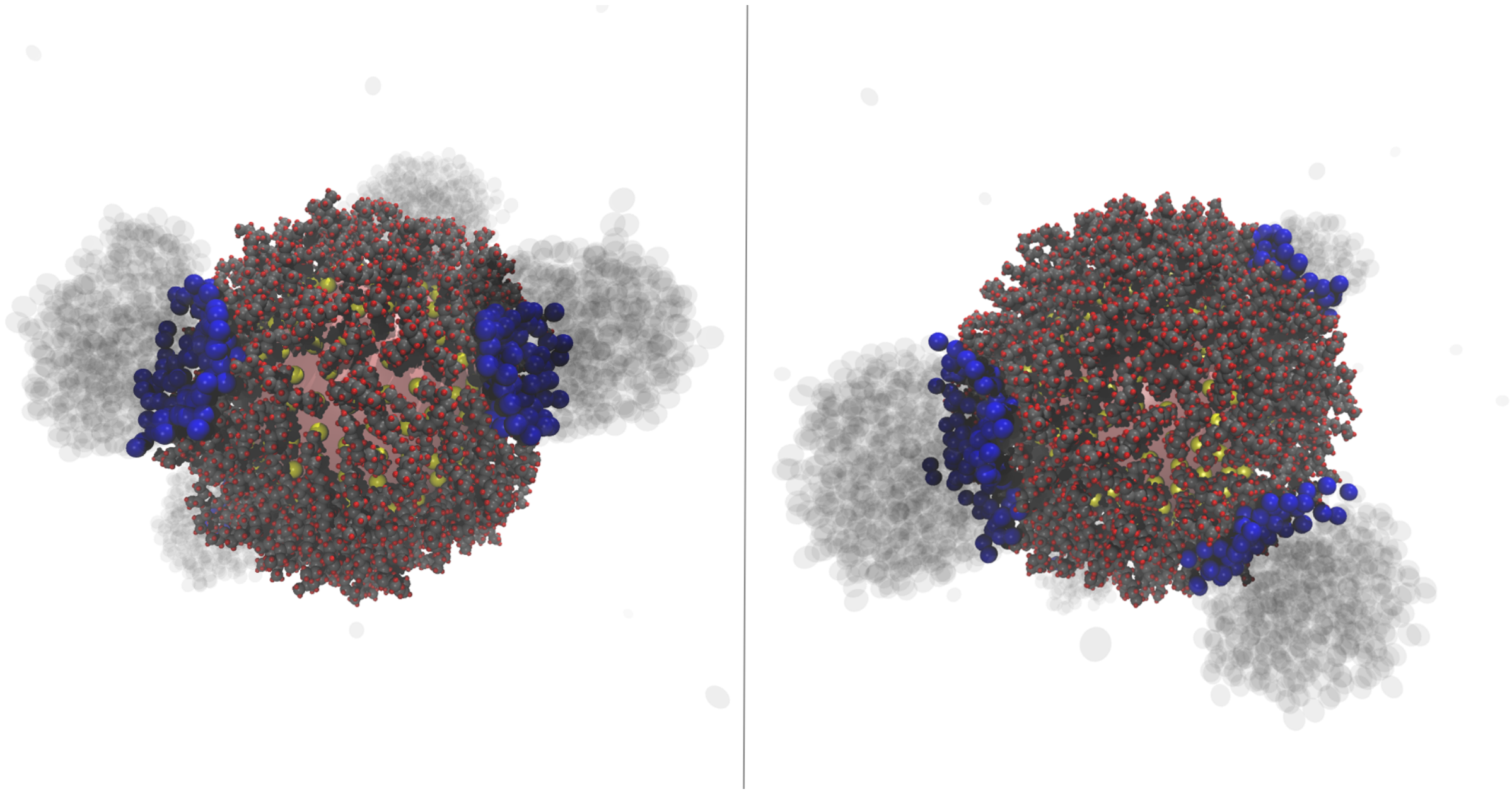

Fig. 3. VMD rendering of an isolated AuNP with water clusters. The layer of water in direct contact with the particle is shown in opaque blue, while the rest of the water cluster is shown in transparent grey. Left panel: 3.6 ligands/nm$^2$. Right panel: 4.5 ligands/nm$^2$. Note the flattening of the water layer and ligand shell in the contact region.

As mentioned above, although both exhibit rapid surface and shape fluctuations, the small water clusters are mobile on the AuNP ligand shell, whereas the large water clusters appear to be almost immobile on the time scale of our simulations. It has been established that water droplet motion can be induced by a strain gradient or a curvature gradient in the supporting surface.[25-29] We suggest that the driving force that generates the mobility of a cluster is ligand shell curvature and strain gradients arising from fluctuations of the chain conformations. To apply this interpretation to the water cluster mobility on a ligated AuNP that we observe, we assume that ligand shell local fluctuations with small spatial extent are more important than fluctuations that span the diameter of the contact area of the cluster because they create larger local gradients, and that these local fluctuations in alkane chain conformations are uncorrelated in space. Then a typical difference in fluctuation induced curvature or strain between opposite sides of a cluster contact area will create a larger gradient across the diameter of a small cluster than across the diameter of a large cluster. Given the mass ratio of the large and small water clusters, it is plausible to expect the mobility of the large adsorbed water cluster to be suppressed relative to that of the small cluster.

There is another consequence of the deformation of a soft substrate induced by contact with an adsorbed liquid droplet. Style et al. have established experimentally and theoretically that there is a deformation of a soft solid near the liquid-solid contact line that inhibits coalescence of droplets.[30] Additionally, Roy et al. and Karpitschka et al. have shown that when adsorbed on a thin elastic solid the interaction between water droplets has a repulsive barrier.[31,32] We argue that these attributes of the interaction between a water droplet and a deformable surface contribute to the persistence of separated water droplets adsorbed on the AuNP ligand shell. Further comments on the water droplet distribution are deferred to Sections IIID and IV.

### IIIC. AuNP-AuNP pair potential of mean force in vacuum

We begin the description of the AuNP-AuNP dry and wet pair potentials of mean force obtained from the calculations reported in this paper with a comparison of the dry interaction calculated herein with that calculated for the same AuNPs by Liepold et al.[18] The interest in this comparison is in examination of the relative importance of the discrete H atoms and CH bonds in determining the AuNP-AuNP interaction. We note that the united atom model used by Liepold et al. combines the thiol group with the first carbon, thereby reducing the ligand chain length by one.[18] That difference between all-atom and united atom models can be accounted for by representing the interaction as a function of the core edge-to-core edge AuNP-AuNP separation. As shown in Fig. 4, our vacuum pair potential of mean force matches well that reported by Liepold et al.,[18] noting that the all-atom interaction is slightly deeper and slightly broader than the united atom interaction. The bars attached at each AuNP-AuNP separation for the all-atom simulations represent the range of the interactions for the sampled orientations of the ligand shells at that separation. That non-zero range of AuNP-AuNP interactions at each separation monitors the deviation from spherical symmetry associated with a distribution of the ligands that only partially covers the AuNP. This deviation arises from the constraint, in both the united atom and all-atom models of the AuNP, that the ligand binding sites on the Au core are fixed. This constraint imposes an inhomogeneity on the ligand distribution and a loss of strict rotational symmetry of the AuNP. Suppression of the rotational motion of the AuNPs will then generate a range of possible interactions between pairs of AuNPs. Liepold et al. studied this variation, using the united atom model, for different ligand coverages of the AuNP.[18] They infer that for ligand coverages larger than $\sim 43\%$ the variations in interaction between the different orientations are small.  In contrast, our all-atom simulations reveal that the variations in energy between the different orientations remain significant at 78% and 96% ligand coverage, which we attribute to steric restrictions imposed on the ligand conformations by the discrete H atoms and CH bonds.  These restrictions render the face-to-face distributions more different than represented by the united atom model, as shown in Fig. 4, although the energy averaged over orientations differs little between the all-atom and united atom models.  Given the origin of the AuNP-AuNP interaction, we expect that range of variation to scale with the change in well depth associated with ligand coverage, specifically, that the range will increase as the coverage increases.

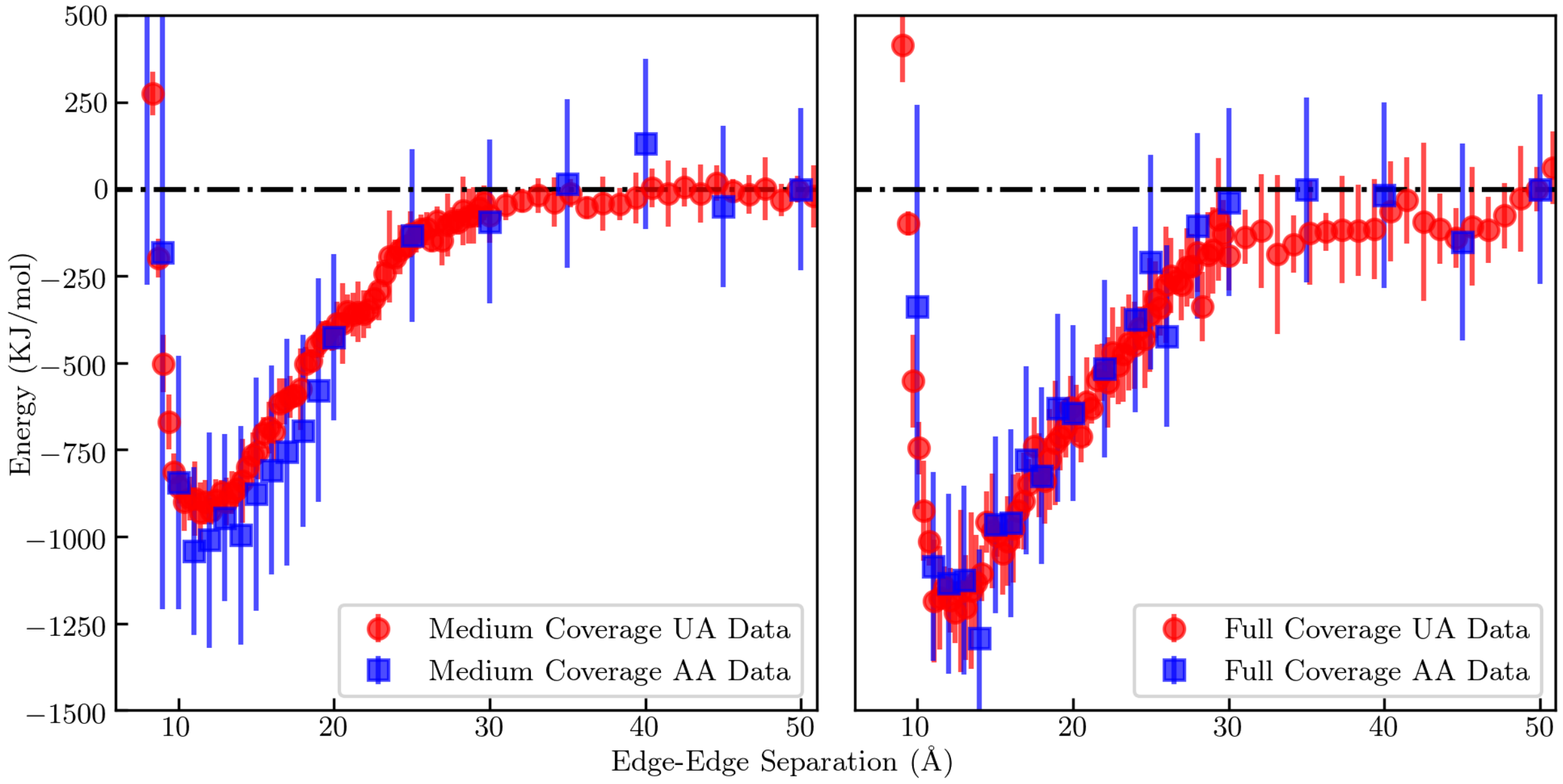

Fig. 4. United-atom (UA) AuNP pair potential of mean force as reported by Liepold et al. compared with the vacuum all-atom (AA) pair potential of mean force obtained in this work: left panel 3.6 ligands/nm$^2$, right panel 4.4 ligands/nm$^2$.[18] The use of the core edge-to-core edge separation rather than the center-center separation allows for model differences between the calculations of the representations of the S terminal group of the ligand and its attachment to the Au core.  The bars along each curve represent the range of interaction generated by different face-to-face orientations of the partially ligated AuNPs.

**IIID. AuNP-AuNP pair potential of mean force in water vapor**

The most obvious results of our simulations of the adsorption of water on a single AuNP and on a pair of AuNPs concern the distribution of water molecules on and between the ligand shells. Consider the AuNPs with coverage of 3.6 ligands/nm$^2$. Fig. 5 displays snapshots of

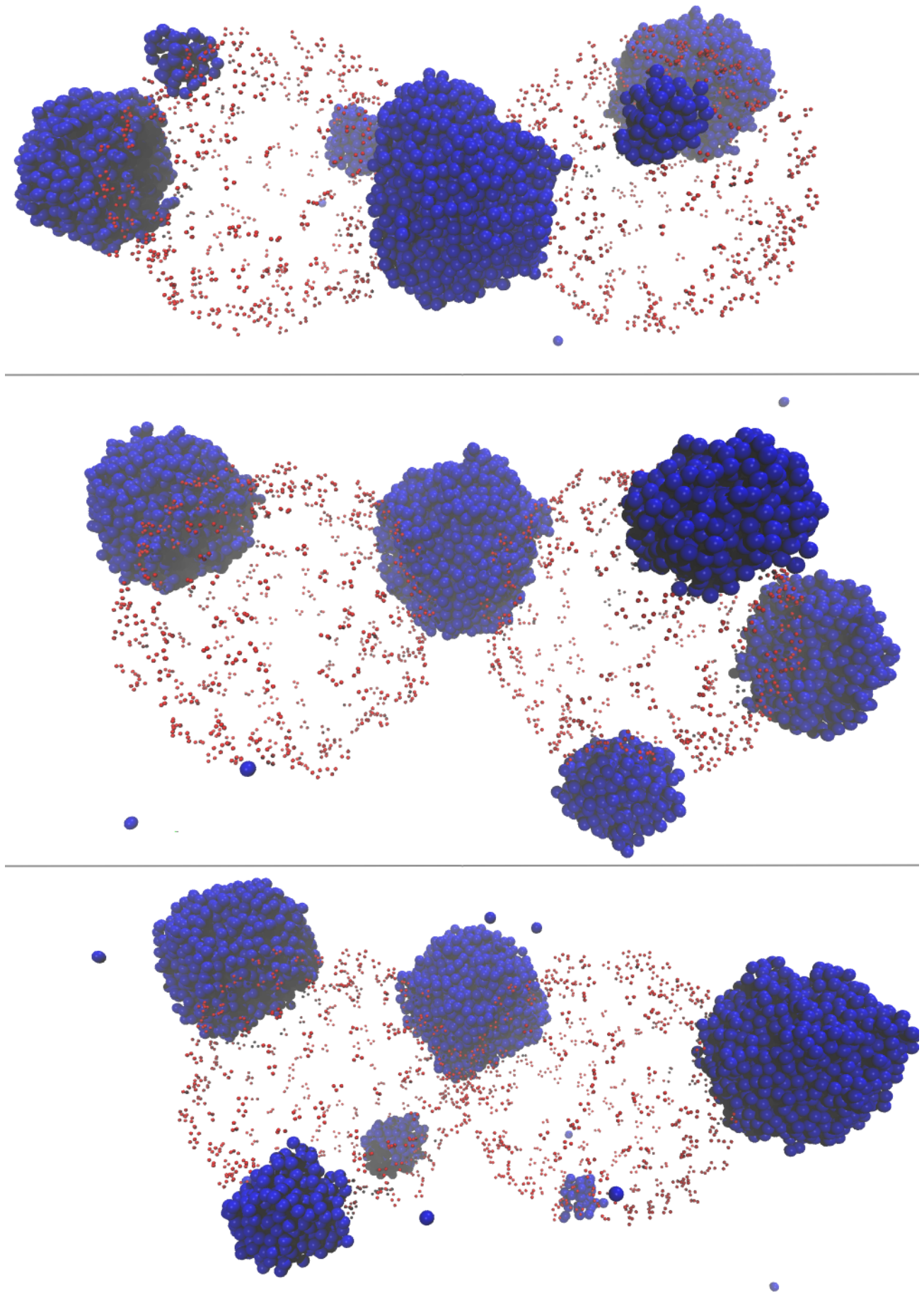

Fig. 5. [One-Column] VMD renderings of a pair of interacting water vapor-exposed, medium-coverage AuNPs at different center-to-center separations (98, 88, and 78 Å from top to bottom). For clarity, only the water and the end group hydrogens are shown.

equilibrium water cluster shapes and distributions on the AuNP ligand shells for several AuNP-AuNP center-to-center separations. As for the single wet AuNP, water clusters form all over both particles. We find that from AuNP-AuNP contact to 100 Å center-to-center separation a water cluster bridges the AuNPs. Different simulation runs started from the initial condition with water molecules uniformly distributed in the simulation cell volume always generate a stationary state with multiple clusters adsorbed on the AuNPs, usually with different numbers of clusters and water molecules per cluster, but in every case with a single cluster bridging the two

AuNPs.  That bridge cluster, independent of its size, typically has a nearly constant number of water molecules over large stretches of AuNP-AuNP separation with that fixed orientation, but there is occasionally a jump in the number of water molecules in the bridge that arises from fusion with another cluster or from fragmentation of the bridge cluster (Fig. 6). An example of the bridge cluster absorbing another cluster is given in Figure 7 and its accompanying video. This jump in number of water molecules is often, but not always, quickly followed by a reversal.

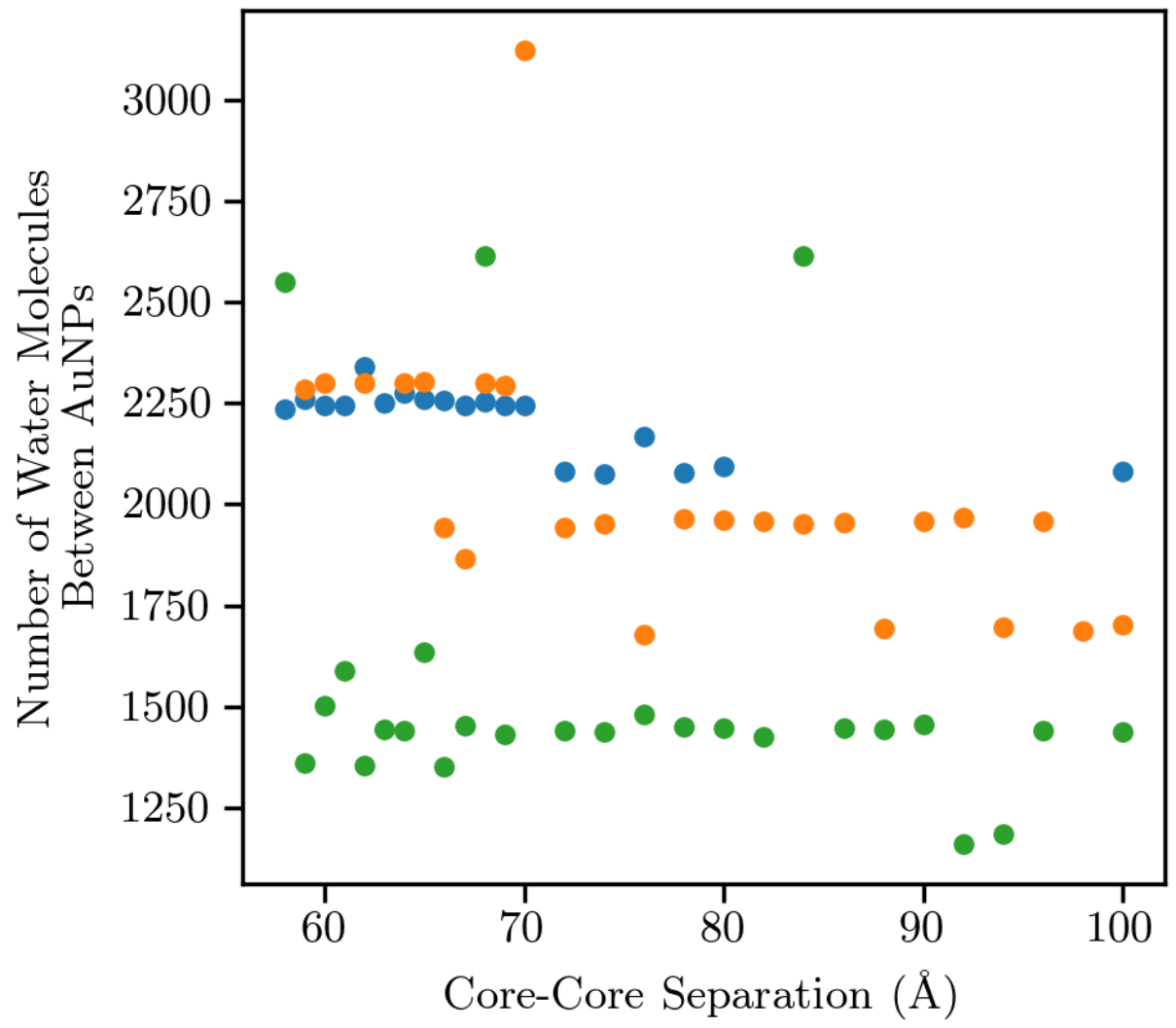

Fig. 6.  [One-Column] Number of water molecules in the cluster bridging two AuNPs as a function of center-to-center separation.  The different color points refer to three AuNP-AuNP face-to-face orientations at medium ligand separation.

The bridge cluster is ubiquitous in all of the AuNP-AuNP orientations sampled; it has the shape of an incomplete collar around the line of centers of the AuNPs, and it flattens the ligand shells of both AuNPs in the regions with which it is in contact.  In the sequence displayed in Fig. 5, at the largest AuNP-AuNP separation the bridge cluster is slightly offset from the line of centers between the particles.  When the AuNP-AuNP separation is continuously decreased to contact the offset increases as the water cluster is forced out by the hydrophobic ligand chains, and once the ligand shells of the two particles touch the bridge cluster sits in the niche formed between the two particles and off the line of centers.  The qualitative picture that emerges has the water bridge remaining adsorbed to both nanoparticles and stretching as the wet AuNP-AuNP separation increases.  For most realizations, the water bridge shape changes with no or very little volume change as the separation changes, noting again the occasional jump in the number of water molecules in the bridge arising from fusion with another cluster or from fragmentation of the bridge cluster, as exemplified by Figure 7. All of these effects are verified by the data displayed in Fig. 6, showing the number of water molecules in the bridge cluster as a function of AuNP-AuNP separation for three fixed AuNP-AuNP ligand shell distributions.

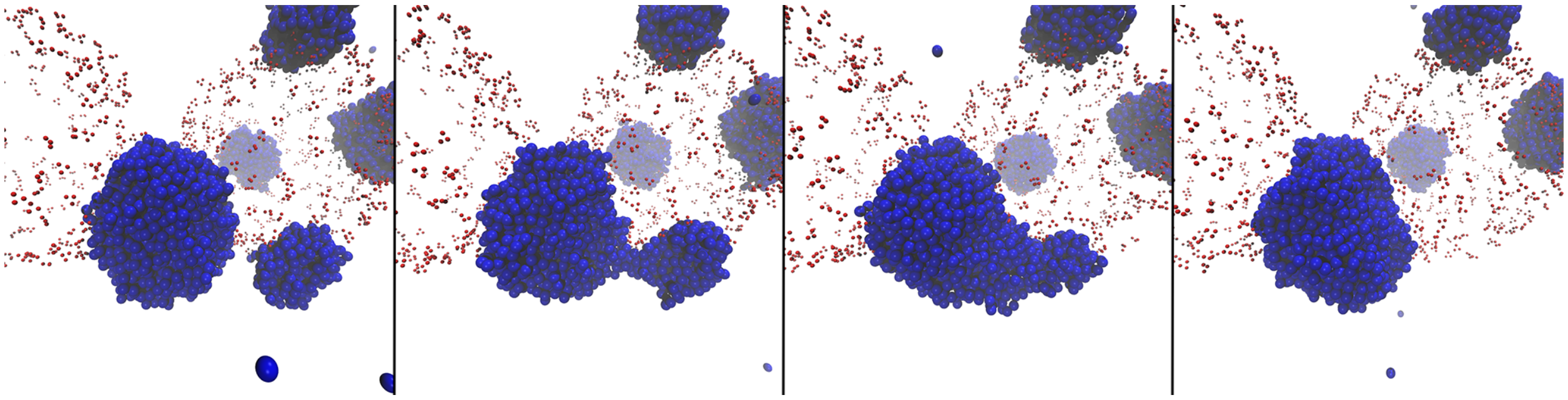

Fig. 7. VMD renderings of a pair of AuNPs at medium ligand coverage in water vapor showing the merger of a water cluster into the bridge cluster. For clarity, only water and end-group hydrogens are shown. The three panels correspond to (left to right) 1.500 ns, 1.525 ns, 1.550 ns, and 1.600 ns into the data-collection phase.

The formation at each separation of a bridge water cluster with sensibly the same number of molecules attests to the stability of the bridge cluster structure. The occasional changes in bridge cluster size are accompanied by changes in shape and in area of contact with the ligand shell, and are associated with fluctuations in interaction energy. We show in Fig. 8 the variation of system energy with number of water molecules in the bridge cluster when the center-to-center AuNP-AuNP separation is 69 Å. The results shown are for five simulations with different face-to-face ligand distributions sampled at times that span the last 0.5 ns of the trajectory in the stationary state reached. The data for the smallest and largest bridges (red points and orange points) display a break in the time signature indicative of clusters merging elsewhere, and the mid-size bridge (green points) displays the signature of a merger with the bridge. Overall, the difference in energy between systems with the smallest and largest bridges is about $\pm 0.3\%$ of the total energy, and the energy fluctuations in any one bridge are about $\pm 0.1\%$ - $\pm 0.2\%$ of the total energy.

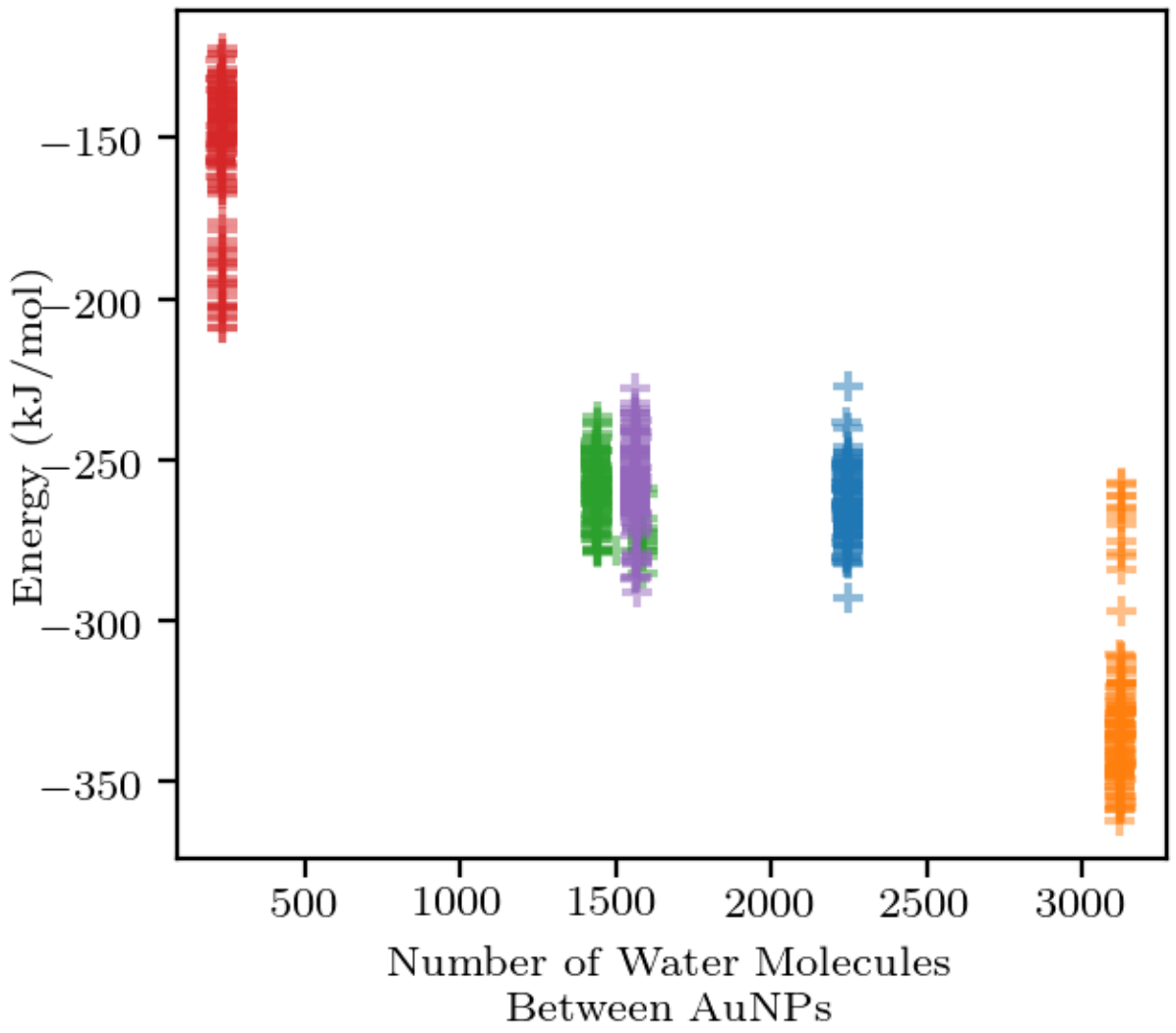

Fig 8. [One-Column] Variation of medium ligand coverage system energy with number of water molecules in the bridge cluster when the AuNP-AuNP center-to-center separation is 69 Å. The data points for each bridge cluster are at intervals that span the last 0.5 ns of the trajectory in the stationary state reached. The energy zero-point is the same as that in Figures 9 and 10.

We display in the left panel of Fig. 9 a comparison of the dry pair potentials of mean force for AuNPs with 78% and 96% ligand coverage. Despite the range of interaction at each separation, the data are consistent with an increase in the well depth by about 20% when the ligand coverage increases from 78% to 96%, in agreement with the well depth change for the same coverage change calculated with the united atom model. We note that the effective diameter of the AuNP is slightly increased when the ligand coverage increases from 78% to 96%, in agreement with the slight extension of the 96% last carbon distribution, relative to the 78% last carbon distribution, displayed in Fig. 1. The range of variation of energy associated with different face-to-face ligand distributions is somewhat increased when the ligand coverage is increased from 78% to 96%, as is expected given the change in well depth.

We display in the right panel of Fig. 9 a comparison of the wet pair potentials of mean force for AuNPs with 78% and 96% ligand coverage. Note that the well depth of the wet pair potential of mean force increases when the ligand coverage increases from 78% to 96%; that increase is about the same as the well depth increase the dry pair potential of mean force with the corresponding change in ligand coverage.

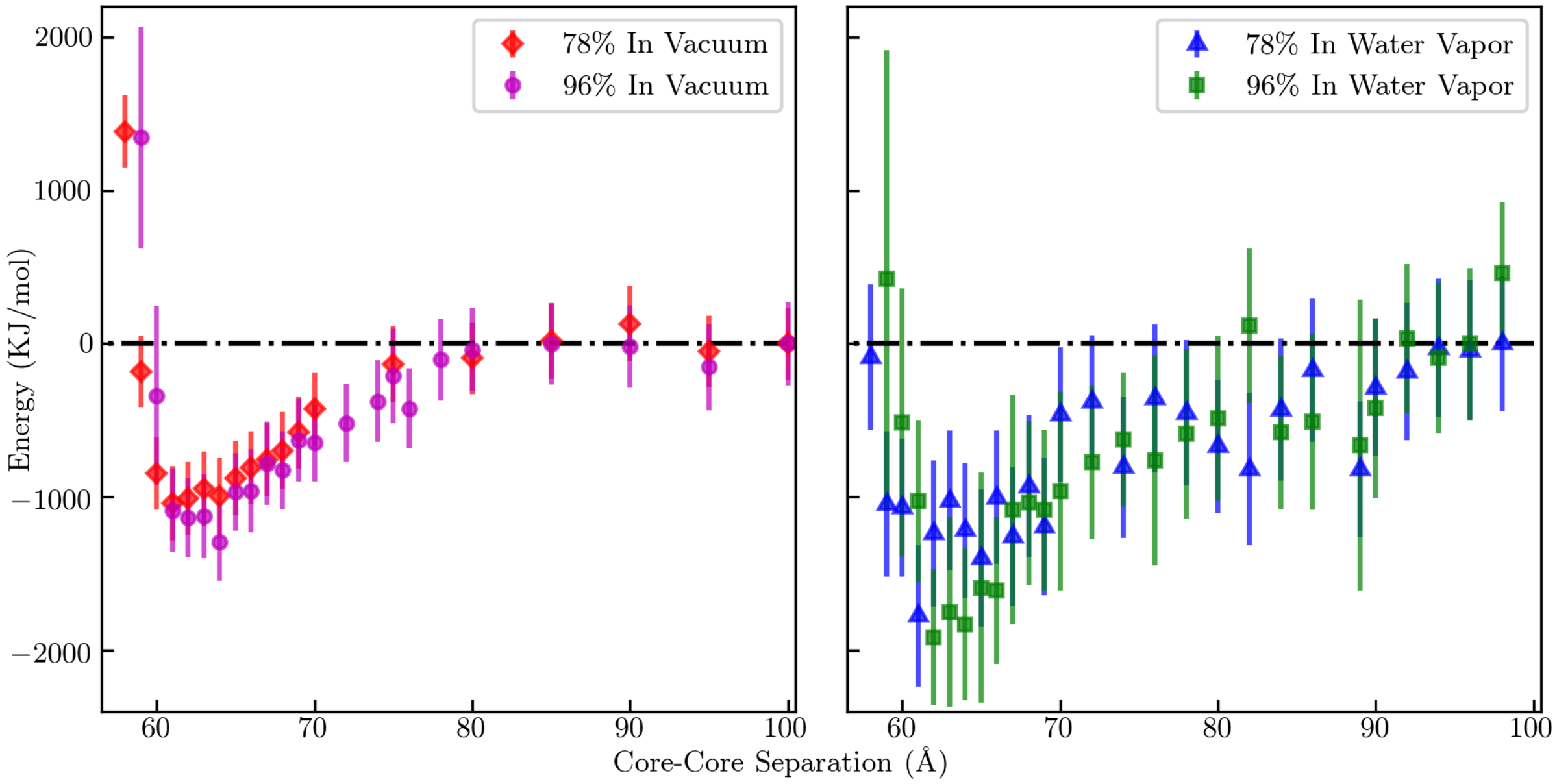

Fig. 9. Comparison between the all atom dry pair potentials of mean force (left panel) and wet pair potentials of mean force (right panel) of AuNPs with 78% and 96% ligand coverage.

As noted earlier, our simulation data show that the water bridge between two AuNPs does not fully surround the line of centers. This finding is confirmed by the results of supplementary simulations that are initiated with all of the water molecules in a liquid ring around the line of centers (see Section IV). These simulations show the initial water slab deforming into a torus between the two nanoparticles. Asymmetry in the torus increases until, at equilibrium, a single large water cluster forms that is between the two nanoparticles but off of the interparticle axis. Although a non-axial water bridge configuration at first sight seems to be inconsistent with the axial symmetry of the pair of AuNPs, Farmer and Bird have shown that minimization of the total surface energy for given volume and contact angle of a liquid bridge between two equal sized spheres generates a shape that does not surround the line of centers if the several liquid-solid contact angles are greater than 90°, which is the case for our system.[33]

A comparison of the dry and wet pair potentials of mean force for the two ligand coverages studied is displayed in Fig. 10. Notwithstanding the fluctuations in the data, the AuNP-AuNP wet pair potential of mean force exhibits qualitative deviation from the dry pair potential of mean force, with a slightly deeper well depth and slightly wider well shape.

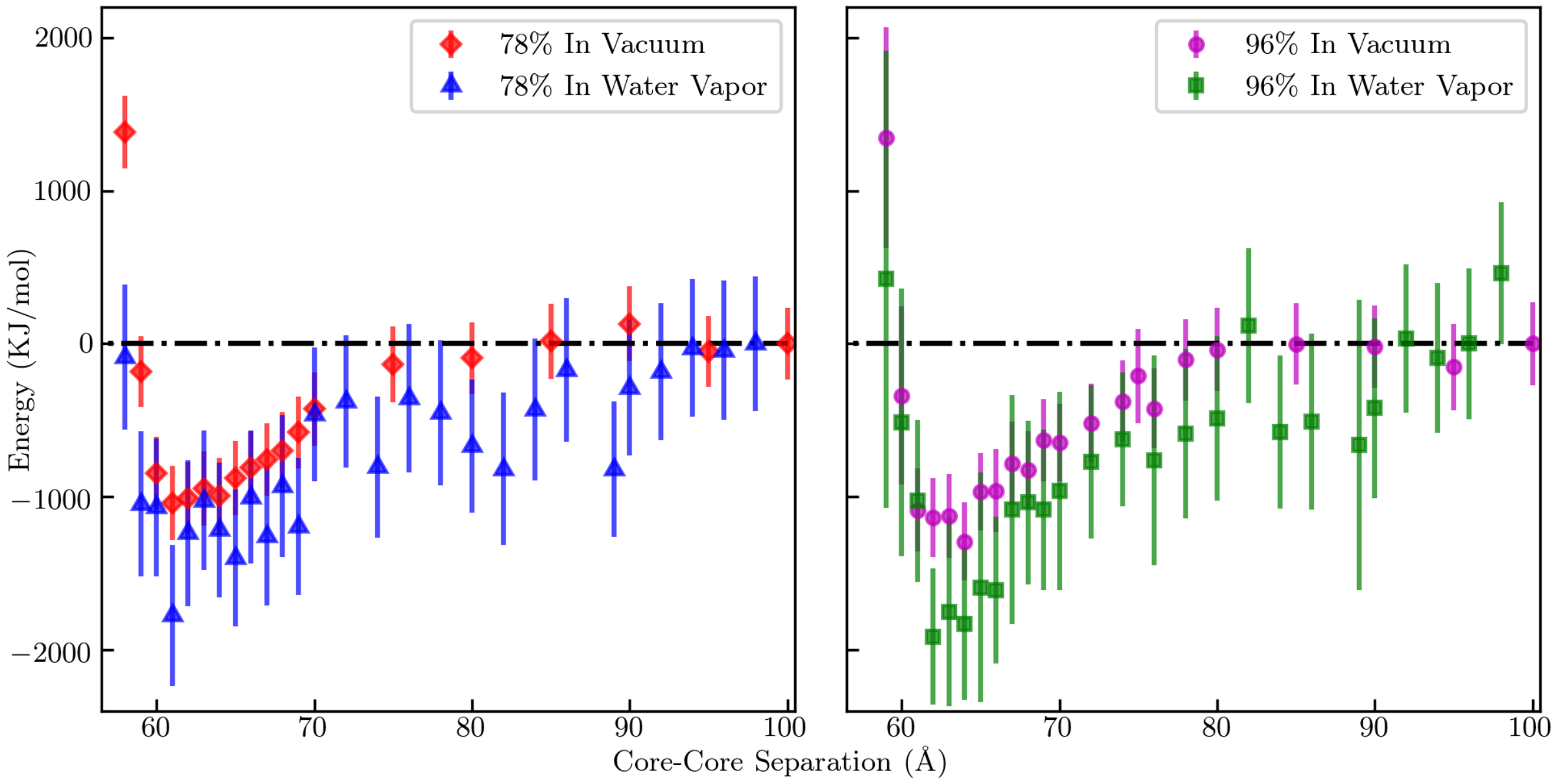

Fig. 10. Comparison of the wet and dry pair potentials of mean force between AuNPs with 78% and 96% ligand coverage.

## IV. Discussion

The stationary state reached from the initial condition with water molecules dispersed throughout the simulation cell volume always has several water clusters, one of which bridges the two AuNPs. In contrast, the stationary state reached from the initial condition with all water molecules in a slab enclosing the line of centers between two AuNPs has only a bridge cluster and no other water clusters (Fig. 11 and its accompanying video). We believe it likely that the one cluster configuration is the equilibrium state for water vapor at 300 K adsorbed on a dodecane thiol ligated AuNP, that the few-cluster configuration is infinitesimally displaced from equilibrium, and that the approach to equilibrium is slowed by the physical characteristics of the water-ligand shell contact described in Section IIIB.

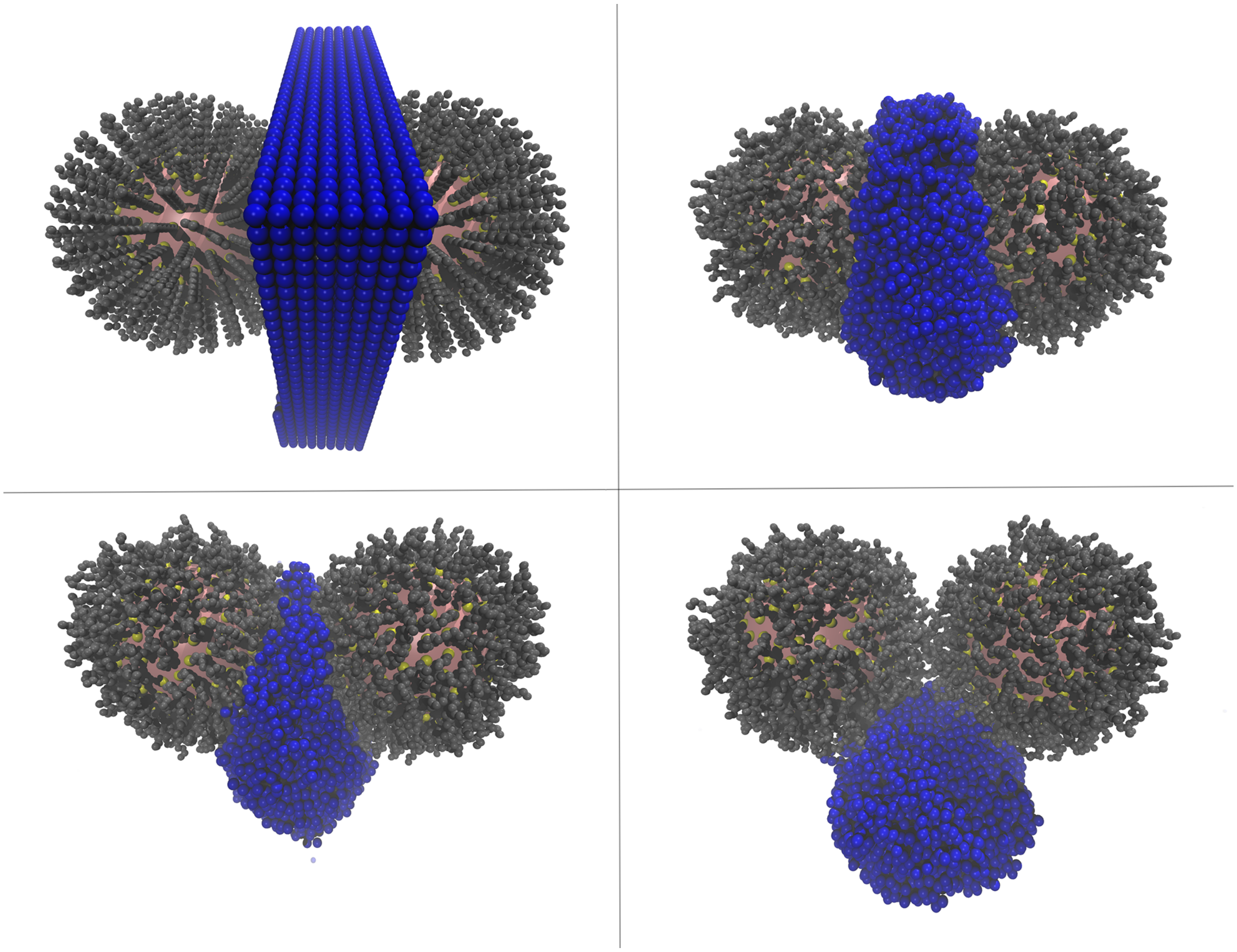

Fig. 11. VMD renditions of a large water cluster adsorbed to the ligand shells of two AuNPs at initialization (top left, 0 ns), equilibration (top right and lower left, 0.4 ns, 0.8 ns), and equilibrium (bottom right, 1.2 ns). The two AuNPs have a coverage of 3.6 ligands/nm$^2$ and a center-to-center separation of 69 Å.

Although we expect that the water clusters adsorbed on the ligand shell of an AuNP are too small to be quantitatively described with only thermodynamic and hydrodynamic considerations, we have found that such analyses of macroscopic water-surface interactions provide both guidance to understanding and qualitatively correct predictions of the properties and phenomena observed in our simulations. Thus, we find that the simulated water cluster-ligand shell and the measured bulk water-planar paraffin surface contact angles are the same, suggesting that surface tension and line tension forces associated with water cluster-ligand shell interaction are well represented by the corresponding macroscopic parameters. Indeed, the flattening of the contact area between water and a deformable surface, the resistance of multiple droplets to forming one large droplet, the formation of a water bridge between AuNPs and the nonaxial symmetry of that bridge, all occur in both macroscopic systems and the simulated water vapor-AuNP system.

Pursuing further insight from macroscopic analyses concerning how the water bridge between the AuNPs generates an adhesive force, we note that in the macroscopic situation interposing a very thin liquid layer between two surfaces generates an adhesive force. That adhesive force depends on surface geometry, liquid-solid contact angle, liquid surface tension, surface roughness and deformation, and surface motion. Macroscopic analyses of this adhesive force assume the liquid bridge is axially symmetric, whereas our simulations show that the water cluster does not encircle the core-core line of centers. Nevertheless, a qualitative picture of how the water bridge between two AuNPs generates an adhesive force, and the dependence of that force on AuNP-AuNP separation, can be obtained from consideration of two model cases: the adhesive force between two rigid flat plates and the adhesive force between two spheres, each generated by a thin liquid film in contact with both surfaces. In both cases the adhesive force is inversely proportional to the square of the separation $H$ of the surfaces.[34] We note that the analytic representation of the force is invalid when this separation becomes comparable with a few water molecule diameters. In that domain we expect that the dominant contribution to the changes in the pair potential of mean force arises from a combination of molecular reorganization in the water immediately surrounding the AuNPs, e.g., the local tetrahedral order in the water, and perturbation of the conformations of the ends of ligands under water clusters. For separations larger than a few water molecule diameters, returning to the macroscopic description, the film sourced adhesive force can be thought of as derived from a long-ranged interaction energy that varies as $-1/H$, which lowers the pair potential of mean force and alters its shape.

The changes in the portion of the ligand shell under a water cluster that we find are like those inferred from Monte Carlo simulations of the full solvation of an isolated ligated AuNP in bulk SPC water reported by Prasad and Gupta.[7] They find that bulk water barely penetrates the ligand shell of the AuNP, consistent with our finding vis-a-vis the interface between a water cluster and the ligand shell. They also show that the effect of the bulk water environment on the shape of the radial density distribution of ligands is very weak, just as is the case for the clusters adsorbed on the AuNP (Fig. 1), and that the reduction of the radius of gyration of the ligand shell of the water immersed AuNP from its value in vacuum is 4%, which is consistent with the small shift in peak density of the distribution of the last carbon in the ligand chain for chains covered by a water cluster (Fig. 1).

Our calculations of the difference between the wet and dry pair potentials of mean force between AuNPs support two qualitative inferences vis a vis the properties of an ordered monolayer of AuNPs. First, we verify that the water vapor induced change in pair potential of mean force is a surrogate indicator of how exposure to water vapor changes the Young's modulus the AuNP monolayer. To do so we note that all lattice models used to calculate the Young's modulus of a 2D system with specified pair interactions, irrespective of lattice symmetry or range and nature of the interactions, show that it is proportional to the curvature of the interaction around its minimum. If the pair potential of mean force dominates the total interaction in the ordered monolayer, the direction in which the curvature at its minimum changes between the dry and wet cases will identify the direction of change of the Young's modulus. Indeed, the qualitative change in curvature of the pair potential of mean force displayed in Fig. 10 implies that the Young's modulus of a wet monolayer of AuNPs is smaller than that of a dry one, as observed in the experiments. Second, if the ligand chain head groups remain fixed on the Au core surface, as we have assumed, and the distribution of ligands on the particle surface deviates from strict spherical symmetry, the AuNP-AuNP interactions in a monolayer that is self-assembled via evaporation of suspending solvent, and in which the AuNPs do not rotate, will have locally heterogeneous energy environments. This behavior will arise if at the end of the evaporation process the contact between AuNPs randomly arrests diffusive rotations, creating thereby different interactions between a central AuNP and each of its nearest neighbors in the monolayer. Whether or not that heterogeneity in local interactions significantly affects the material properties of the monolayer remains to be established.

## V. Acknowledgements

We thank Miaochen Jin and Kevin B. Slater for assistance with computations in the early stage of the research the reported herein. This research was mainly funded by the University of Chicago Materials Research Science and Engineering Center via the National Science Foundation (DMR-1420709). Binhua Lin also acknowledges financial support from NSF-ChemMatCARS grant (NSF/CHE-1834750). We have used a modified (unpublished) version of DASH provided by D. Reid, J. Helfferich, M. Webb, B. Keene, P. Rauscher, S. Wyetzner and J. de Pablo.

## VI. Data Availability

The data that support the findings of this study are available from the corresponding author upon reasonable request.